\documentclass[preprint,12pt]{elsarticle}

\usepackage{amssymb}
\usepackage{multirow}
\usepackage{booktabs}
\usepackage{amssymb}
\usepackage{mathrsfs}
\usepackage{longtable}
\usepackage{lscape}
\usepackage{tabularx}
\usepackage{epsfig}
\usepackage{colortbl}
\usepackage{tabularx}
\usepackage{subfigure}
\usepackage[mathlines]{lineno}
\biboptions{compress}

\journal{xx}

\begin{document}

\begin{frontmatter}



\title{Quantum games of opinion formation based on the Marinatto-Weber quantum game scheme}


\author[swu,CQS]{Xinyang Deng}
\author[swu,vu]{Yong Deng\corref{cor}}
\author[CQS,DBI]{Qi Liu}
\author[swu,Japan]{Zhen Wang\corref{cor}}

\cortext[cor]{Corresponding authors: ydeng@swu.edu.cn (Y. Deng), zhenwang0@gmail.com (Z. Wang). School of Computer and Information Science, Southwest University, Chongqing 400715, China.}

\address[swu]{School of Computer and Information Science, Southwest University, Chongqing, 400715, China}
\address[CQS]{Center for Quantitative Sciences, Vanderbilt University School of Medicine, Nashville, TN, 37232, USA}
\address[vu]{School of Engineering, Vanderbilt University, Nashville, TN, 37235, USA}
\address[DBI]{Department of Biomedical Informatics, Vanderbilt University School of Medicine, Nashville, TN, 37232, USA}
\address[Japan]{Interdisciplinary Graduate School of Engineering Sciences, Kyushu University, Kasuga-koen, Kasuga-shi, Fukuoka 816-8580, Japan}

\begin{abstract}
Quantization becomes a new way to study classical game theory since quantum strategies and quantum games have been proposed. In previous studies, many typical game models, such as prisoner's dilemma, battle of the sexes, Hawk-Dove game, have been investigated by using quantization approaches. In this paper, several game models of opinion formations have been quantized based on the Marinatto-Weber quantum game scheme, a frequently used scheme to convert classical games to quantum versions. Our results show that the quantization can change fascinatingly the properties of some classical opinion formation game models so as to generate win-win outcomes.
\end{abstract}
\end{frontmatter}

\section{Introduction}
In realistic life, there are large amount of opinion interactions on many issues of interest among social individuals. Opinion formation and evolutionary dynamics have become very interesting and meaningful research fields \cite{gale2003bayesian,castellano2009statistical,acemoglu2011opinion}. Many models, such as DeGroot model \cite{degroot1974reaching}, Sznajd model \cite{sznajd2000opinion}, discrete CODA model \cite{martins2008continuous}, Hegselmann-Krause model \cite{hegselmann2002opinion}, generalized Glauber models \cite{castellano2011irrelevance}, Deffuant model \cite{deffuant2000mixing}, have been developed to simulate interaction mechanisms of opinions. In recent years, game theory, which provides a useful framework to build foundational models to mimick the interactions among agents with conflict of interests in many disciplines \cite{nowak2006five,szabo2015congestion,wang2015evolutionaryEPJB,szolnoki2015conformity,wang2014degree,gracia2014intergroup,deng2014impact,santos2005scale,wang2015ScalingPLR,boccaletti2014structure}, has also been used in these fields to help understand opinion formation and evolution. With respect to opinion formation, a lot of game models have been proposed or used in previous studies \cite{di2007opinion,ding2010evolutionary,ding2009co,cao2008mixed,zheng2015bargaining,gargiulo2012influence}. Among these, several models given by Di Mare and Latora \cite{di2007opinion} has attracted much interest due to the simplicity and affluent meanings of those models.

Quantum game is a new breakthrough of game theory, inspired by quantum information theory. It extends the classical strategy space to a quantum strategy space, and rules players playing the game by following quantum rules. At present, there are several typical quantum game schemes, for example Eisert-Wilkens-Lewenstein scheme \cite{eisert1999quantum} and Marinatto-Weber scheme \cite{marinatto2000quantum}, to quantize the classical strategy space so as to build quantum games. Quantization approaches have already been adopted to many classical game models, for example prisoner's dilemma \cite{eisert1999quantum,li2014entanglement}, battle of the sexes \cite{marinatto2000quantum}, Hawk-Dove game \cite{nawaz2010evolutionarily,hanauske2010doves}, and so on \cite{eisert2000quantum,du2002entanglement,benjamin2001multiplayer,sharif2013introduction,flitney2002quantum,guo2008survey,frkackiewicz2011application,situ2014quantum}. Through quantization, these game models show some different characteristics, compared with their classical counterparts. However, the game models of opinion formation are not yet studied by using quantization approaches. It is very interesting to see what happens and whether new features emerge if quantizing the classical opinion formation game models. Motivated by this idea, the quantization of opinion formation games are studied in this paper. During this study, we follow and use Di Mare and Latora's models to simulate the interactions between two individuals with different opinions, and Marinatto-Weber scheme to quantizing the classical game models. The results show that the quantization can change fascinatingly the properties of some classical opinion formation game models so as to generate win-win outcomes.

\section{Preliminaries}
\subsection{Game models of opinion formation}
In Di Mare and Latora's game models of opinion formation \cite{di2007opinion}, two players, row player $A$ and column player $B$, paly one-shot games with three strategies, namely to keep one's own opinion (\emph{Keep}), to change one's own opinion (\emph{Change}), and to take a compromise opinion (\emph{Agree}). When players take different actions, they receive different payoffs:
\begin{itemize}
  \item $s = +a$ if the other player changes her opinion
  \item $s = +b$ if the player keeps her opinion
  \item $s = -b$ if the other player keeps her opinion
  \item $s = -a$ if the player changes her opinion
  \item $s = +c$ or $s = +c + 1/d$ if the other player changes her opinion with an intermediate one
  \item $s = -c$ or $s = -c + 1/d$ if the player changes her opinion with an intermediate one
\end{itemize}
where $a,b,c,d > 0$. Based on these payoffs mentioned above, three different payoff matrices are derived.

Game model (GM) I:
\begin{equation}\label{PayoffMatrixGMI}
\begin{array}{*{20}c}
   {} & {\begin{array}{*{20}c}
   {Chang} & {\quad \quad \quad Keep}  \\
\end{array}}  \\
   {\begin{array}{*{20}c}
   {Chang}  \\
   {Keep}  \\
\end{array}} & {\left( {\begin{array}{*{20}c}
   {(0,0)} & {( - a - b, + a + b)}  \\
   {( + a + b, - a - b)} & {(0,0)}  \\
\end{array}} \right)}  \\
\end{array}
\end{equation}

GM II:
\begin{equation}\label{PayoffMatrixGMII}
\begin{array}{*{20}c}
   {} & {\begin{array}{*{20}c}
   {Chang} & {\quad \quad \quad Keep\quad } & {\quad \qquad Agree}  \\
\end{array}}  \\
   {\begin{array}{*{20}c}
   {Chang}  \\
   {Keep}  \\
   {Agree}  \\
\end{array}} & {\left( {\begin{array}{*{20}c}
   {(0,0)} & {( - a - b, + a + b)} & {( - a + c, + a - c)}  \\
   {( + a + b, - a - b)} & {(0,0)} & {( + b + c, - b - c)}  \\
   {( + a - c, - a + c)} & {( - b - c, + b + c)} & {(0,0)}  \\
\end{array}} \right)}  \\
\end{array}
\end{equation}

GM III:
\begin{equation}\label{PayoffMatrixGMIII}
\begin{array}{*{20}c}
   {} & {\begin{array}{*{20}c}
   {Chang} & {\qquad \qquad \qquad \qquad Keep \qquad } & {\qquad \qquad \qquad Agree}  \\
\end{array}}  \\
   {\begin{array}{*{20}c}
   {Chang}  \\
   {Keep}  \\
   {Agree}  \\
\end{array}} & {\left( {\begin{array}{*{20}c}
   {(0,0)} & {( - a - b, + a + b)} & {( - a + c + \frac{1}{d}, + a - c + \frac{1}{d})}  \\
   {( + a + b, - a - b)} & {(0,0)} & {( + b + c + \frac{1}{d}, - b - c + \frac{1}{d})}  \\
   {( + a - c + \frac{1}{d}, - a + c + \frac{1}{d})} & {( - b - c + \frac{1}{d}, + b + c + \frac{1}{d})} & {(\frac{2}{d},\frac{2}{d})}  \\
\end{array}} \right)}  \\
\end{array}
\end{equation}

GM I, which does not consider the strategy ``Agree", is a zero-sum game, and it has a single Nash equilibrium (NE) $(Keep, Keep)$. GM II is also a zero-sum game whose unique NE is $(Keep, Keep)$ as well. GM III contains some differences: Firstly, GM III is not a zero-sum game; Secondly, different from GM I and GM II, in GM III the distance between two players' opinions, denoted as $d$, has been taken into consideration based on such idea that two individuals are much easier to reach an agreement if their opinions are closer; Third, GM III has two different NE points corresponding to different conditions, one is $(Keep, Keep)$ where the joint payoff of two players is 0 if $d \ge 1/(b+c)$, the other is $(Agree, Agree)$ where the joint payoff is $4/d$ if $d \le 1/(b+c)$.

It is easy to prove that the point $(Agree, Agree)$ is the Pareto optimality of GM III. At that point, the outcome of the game between these two players with different opinions is not a zero-sum, but a win-win. However, in GM III it is conditional to realize the Pareto optimality. If $d > 1/(b+c)$, GM III still generates zero-sum outcomes, as same as GM I and GM II.

\subsection{MW scheme for strategic game}
Marinatto-Weber (MW) scheme \cite{marinatto2000quantum} is one of the most frequently used quantum game schemes, which is developed for quantizing matrix games with $2 \times 2$ dimension. Here, we use the battle of the sexes as an example to introduce the MW scheme.

In the game of battle of the sexes, players are Alice and Bob, each of them has two choices, opera ($O$) and football ($F$). The payoffs are different for various strategy combinations of Alice and Bob, for example $E_A (O,F)$ represents Alice's payoff when she chooses strategy $O$ and Bob chooses strategy $F$. Then, a four-dimensional Hilbert space $H$ can be defined for the battle of the sexes by giving its orthonormal basis vectors $H = H_A \otimes H_B = \{ \left| {OO} \right\rangle, \left| {OF} \right\rangle, \left| {FO} \right\rangle, \left| {FF} \right\rangle \}$, where the first qubit is reserved to the state of Alice and the second one to that of Bob.

In the initial, assuming that Alice and Bob share the following quantum state:
\begin{equation}\label{PsiInABCD}
\left| {\psi _{in} } \right\rangle  = u_{11}\left| {OO} \right\rangle  + u_{12}\left| {OF} \right\rangle  + u_{21}\left| {FO} \right\rangle  + u_{22}\left| {FF} \right\rangle
\end{equation}
where $|u_{11}|^2 + |u_{12}|^2 + |u_{21}|^2 + |u_{22}|^2 = 1$. According to state vector $\left| {\psi _{in} } \right\rangle$, the associated density matrix can be derived as $\rho _{in}  = \left| {\psi _{in} } \right\rangle \left\langle {\psi _{in} } \right|$.

Let $C$ be a unitary and Hermitian operator (i.e., $C^\dag   = C = C^{ - 1}$) such that $C\left| O \right\rangle = \left| F \right\rangle$, $C\left| F \right\rangle = \left| O \right\rangle$, and $I$ be the identity operator. In the game process, Alice and Bob use operators $I$ and $C$ with probabilities $p$, $(1-p)$, $q$, $(1-q)$, respectively. Then, the final density matrix for this two-qubit quantum system takes the form:
\begin{equation}\label{RhoFinal}
\begin{array}{l}
\rho_{fin}   = pq\left[ {(I_A  \otimes I_B )\rho _{in} (I_A^\dag   \otimes I_B^\dag  )} \right] \\
 \quad \quad \quad  + p(1 - q)\left[ {(I_A  \otimes C_B )\rho _{in} (I_A^\dag   \otimes C_B^\dag  )} \right] \\
 \quad \quad \quad  + (1 - p)q\left[ {(C_A  \otimes I_B )\rho _{in} (C_A^\dag   \otimes I_B^\dag  )} \right] \\
 \quad \quad \quad  + (1 - p)(1 - q)\left[ {(C_A  \otimes C_B )\rho _{in} (C_A^\dag   \otimes C_B^\dag  )} \right] \\
 \end{array}
\end{equation}

In order to calculate the payoffs, two payoff operators are introduced:
\begin{equation}\label{PayoffOperatorPA}
\begin{array}{l}
 P_A  = E_A (O,O)\left| {OO} \right\rangle \left\langle {OO} \right| + E_A (O,F)\left| {OF} \right\rangle \left\langle {OF} \right| \\
 \quad \quad  + E_A (F,O)\left| {FO} \right\rangle \left\langle {FO} \right| + E_A (F,F)\left| {FF} \right\rangle \left\langle {FF} \right| \\
 \end{array}
\end{equation}
\begin{equation}\label{PayoffOperatorPB}
\begin{array}{l}
 P_B  = E_B (O,O)\left| {OO} \right\rangle \left\langle {OO} \right| + E_B (O,F)\left| {OF} \right\rangle \left\langle {OF} \right| \\
 \quad \quad  + E_B (F,O)\left| {FO} \right\rangle \left\langle {FO} \right| + E_B (F,F)\left| {FF} \right\rangle \left\langle {FF} \right| \\
 \end{array}
\end{equation}

Finally, the payoff functions of Alice and Bob can be obtained as mean values of these operators:
\begin{equation}\label{PayoffFunctionsTrAB}
\bar \$ _A (p,q) = Tr(P_A \rho _{fin} ),\quad \bar \$ _B (p,q) = Tr(P_B \rho _{fin} ).
\end{equation}

As stated above, (\ref{PsiInABCD}) - (\ref{PayoffFunctionsTrAB}) clearly show how the MW scheme quantizes a classical game. It is worthy noting that the original MW scheme is just able to deal with $2 \times 2$ matrix games. Several works \cite{iqbal2002quantum,iqbal2004stability,frkackiewicz2013new} are devoted to study the generalization of classical MW scheme, which strengthens the ability of this quantum scheme.

\section{Quantizing opinion formation games via the MW scheme}
In this paper, we follow the MW scheme to quantize the classical game models of opinion formation, GM I, GM II, and GM III. In terms of these quantum game models, we investigate the impact of quantization on the classical opinion formation games. Especially, whether the joint payoff of two players is increased or not by the quantization. In these quantum game models, strategy ``Change'', ``Keep'' and ``Agree'' are simply denoted as $1$, $2$, and $3$, respectively.

\subsection{Quantum GM I}
Let us first quantize the GM I whose payoff matrix is shown as (\ref{PayoffMatrixGMI}). Without loss of generality, initially assuming that two individuals, $A$ and $B$ share the following entangled state:
\begin{equation}
\left| {\psi _{in} } \right\rangle  = u_{11} \left| {11} \right\rangle  + u_{12} \left| {12} \right\rangle  + u_{21} \left| {21} \right\rangle  + u_{22} \left| {22} \right\rangle
\end{equation}
where $\left| {u_{11} } \right|^2  + \left| {u_{12} } \right|^2  + \left| {u_{21} } \right|^2  + \left| {u_{22} } \right|^2  = 1 $., and the first qubit is reserved for strategy of row player $A$ and the second for column player $B$.

Let $C$ be a unitary (i.e., $C^\dag   = C = C^{ - 1}$) such that $C\left| 1 \right\rangle = \left| 2 \right\rangle$, $C\left| 2 \right\rangle = \left| 1 \right\rangle$. Player $A$ performs identity operator $I$ with probability $p$, and $C$ with probability $1-p$. Analogously, player $B$ uses $I$ with probability $q$ and $C$ with probability $1-q$. Then, the final density matrix for this two-qubit quantum system, $\rho _{fin}$, can be obtained according to (\ref{RhoFinal}). Finally, according to (\ref{PayoffOperatorPA}) - (\ref{PayoffFunctionsTrAB}), the expected payoff functions for player $A$ and $B$ can be derived as follow:
\begin{equation}
\begin{array}{l}
 \bar \$ _A (p,q) =  - (a + b)(p|u_{11} |^2  + p|u_{12} |^2  - p|u_{21} |^2  - p|u_{22} |^2  \\
 \quad \quad \quad \quad  - q|u_{11} |^2  + q|u_{12} |^2  - q|u_{21} |^2  + q|u_{22} |^2  - |u_{12} |^2  + |u_{21} |^2 ) \\
 \bar \$ _B (p,q) = (a + b)(p|u_{11} |^2  + p|u_{12} |^2  - p|u_{21} |^2  - p|u_{22} |^2  \\
 \quad \quad \quad \quad  - q|u_{11} |^2  + q|u_{12} |^2  - q|u_{21} |^2  + q|u_{22} |^2  - |u_{12} |^2  + |u_{21} |^2 ) \\
 \end{array}
\end{equation}
It is found that $\bar \$ _A (p,q) + \bar \$ _B (p,q) = 0$, which means that the quantization does not change the zero-sum essence of GM I. For the quantum GM I, a NE $(p^*, q^*)$ can be found by imposing the following two conditions:
\begin{equation}\label{EqNEconditionsGMI}
\begin{array}{l}
 \bar \$ _A (p^* ,q^* ) - \bar \$ _A (p,q^* ) = (a + b)(p - p^* )(|u_{11} |^2  + |u_{12} |^2  - |u_{21} |^2  - |u_{22}|^2 ) \ge 0,\quad \forall p \in [0,1] \\
 \bar \$ _B (p^* ,q^* ) - \bar \$ _B (p^* ,q) = (a + b)(q - q^* )(|u_{11} |^2  - |u_{12} |^2  + |u_{21} |^2  - |u_{22}|^2 ) \ge 0,\quad \forall q \in [0,1] \\
 \end{array}
\end{equation}

From inequalities (\ref{EqNEconditionsGMI}), it is easy to prove that the quantum GM I can be reduced to the classical GM I if there does exist a term which is equal to 1 among $\left| {u_{11} } \right|^2$, $\left| {u_{12} } \right|^2$, $\left| {u_{21} } \right|^2$, $\left| {u_{22} } \right|^2$. Again, due to $\bar \$ _A (p,q) + \bar \$ _B (p,q) = 0$, the quantum GM I is still a zero-sum game, unconditionally. Additionally, there may exist other properties, but they are ignored in this paper since we pay main attention on the trait of joint payoff.

\subsection{Quantum GM II}
Now, let us consider GM II whose payoff matrix is shown as (\ref{PayoffMatrixGMII}). It is noticed that GM II is a $3 \times 3$ game and the original MW quantum scheme is only suitable for $2 \times 2$ games. In \cite{iqbal2002quantum,iqbal2004stability}, Iqbal and Toor extended the classical MW scheme to bi-matrix games with $3 \times 3$ dimension. In this paper, we follow the generalized MW scheme proposed by Iqbal and Toor to build quantum GM II. Similaryly, without loss of generality, the initial state shared by two players, $A$ and $B$, are assumed as follows.
\begin{equation}\label{GMWSchemePsiIn}
\left| {\psi _{in} } \right\rangle  = \sum\limits_{i,j = 1,2,3} {u_{ij} \left| {ij} \right\rangle } \;,\quad \sum\limits_{i,j = 1,2,3} {\left| {u_{ij} } \right|^2 }  = 1.
\end{equation}

In the quantum GM II, each player possesses three unitary operators, $I$, $C$ and $D$, such that
\begin{equation}
\begin{array}{l}
 I\left| 1 \right\rangle  = \left| 1 \right\rangle ,\quad C\left| 1 \right\rangle  = \left| 3 \right\rangle ,\quad D\left| 1 \right\rangle  = \left| 2 \right\rangle , \\
 I\left| 2 \right\rangle  = \left| 2 \right\rangle ,\quad C\left| 2 \right\rangle  = \left| 2 \right\rangle ,\quad D\left| 2 \right\rangle  = \left| 1 \right\rangle , \\
 I\left| 3 \right\rangle  = \left| 3 \right\rangle ,\quad C\left| 3 \right\rangle  = \left| 1 \right\rangle ,\quad D\left| 3 \right\rangle  = \left| 3 \right\rangle . \\
 \end{array}
\end{equation}
where $C^\dag   = C = C^{ - 1}$, $D^\dag   = D = D^{ - 1}$, and $I$ is the identity operator. Suppose player $A$ uses $C$, $D$, $I$ with probabilities $p$, $p_1$, $(1-p-p_1)$. Similarly, player $B$ performs $C$, $D$ and $I$ with probabilities $q$, $q_1$ and $(1-q-q_1)$, respectively. The final density matrix after $A$ and $B$ have played their strategies can be obtained as below.
\begin{equation}
\begin{array}{l}
 \rho _{fin}  = (1 - p - p_1 )(1 - q - q_1 )\left[ {(I_A  \otimes I_B )\rho _{in} (I_A^\dag   \otimes I_B^\dag  )} \right] \\
 \quad \quad  + p(1 - q - q_1 )\left[ {(C_A  \otimes I_B )\rho _{in} (C_A^\dag   \otimes I_B^\dag  )} \right] + p_1 (1 - q - q_1 )\left[ {(D_A  \otimes I_B )\rho _{in} (D_A^\dag   \otimes I_B^\dag  )} \right] \\
 \quad \quad  + (1 - p - p_1 )q\left[ {(I_A  \otimes C_B )\rho _{in} (I_A^\dag   \otimes C_B^\dag  )} \right] + (1 - p - p_1 )q_1 \left[ {(I_A  \otimes D_B )\rho _{in} (I_A^\dag   \otimes D_B^\dag  )} \right] \\
 \quad \quad  + pq\left[ {(C_A  \otimes C_B )\rho _{in} (C_A^\dag   \otimes C_B^\dag  )} \right] + pq_1 \left[ {(C_A  \otimes D_B )\rho _{in} (C_A^\dag   \otimes D_B^\dag  )} \right] \\
 \quad \quad  + p_1 q\left[ {(D_A  \otimes C_B )\rho _{in} (D_A^\dag   \otimes C_B^\dag  )} \right] + p_1 q_1 \left[ {(D_A  \otimes D_B )\rho _{in} (D_A^\dag   \otimes D_B^\dag  )} \right] \\
 \end{array}
\end{equation}
where $\rho _{in}  = \left| {\psi _{in} } \right\rangle \left\langle {\psi _{in} } \right|$. In this situation, assuming that $E_X (i,j)$ represents the payoff of player $X$ when X's strategy is $i$ and the other's strategy is $j$, the payoff operators for $A$ and $B$ are
\begin{equation}
P_A  = \sum\limits_{i,j = 1,2,3} {E_A (i,j)\left| {ij} \right\rangle \left\langle {ji} \right|} \;,\quad P_B  = \sum\limits_{i,j = 1,2,3} {E_B (i,j)\left| {ij} \right\rangle \left\langle {ji} \right|}.
\end{equation}
Finally, the payoffs of $A$ and $B$ can be obtained as follows:
\begin{equation}\label{GMWSchemePayoffFunctions}
\bar \$ _A (p, p_1, q, q_1) = Tr(P_A \rho _{fin} ),\quad \bar \$ _B (p, p_1, q, q_1) = Tr(P_B \rho _{fin} ).
\end{equation}

By means of the generalized MW scheme described above, the GM II can be quantized. Various NEs can be found by changing the initial state $\left| {\psi _{in} } \right\rangle$. The expected payoff functions of $A$ and $B$ are denoted as $\bar \$ _A (p, p_1, q, q_1)$ and $\bar \$ _B (p, p_1, q, q_1)$, respectively. Since the expressions are very long, here we do not give. As same as the quantum GM I, a result is found that $\bar \$ _A (p, p_1, q, q_1) + \bar \$ _B (p, p_1, q, q_1) = 0$, which implies that the quantum GM II is still a zero-sum game. In summary, although the quantization can produce more NEs, the outcomes of quantum GM I and quantum GM II are still zero-sum.

\subsection{Quantum GM III}
Now, let us investigate the GM III whose payoff matrix is given as (\ref{PayoffMatrixGMIII}). By using the generalized MW scheme as shown in (\ref{GMWSchemePsiIn}) - (\ref{GMWSchemePayoffFunctions}), the expected payoff functions $\bar \$ _A (p, p_1, q, q_1)$ and $\bar \$ _B (p, p_1, q, q_1)$ can be derived. Here, our attention has mainly been paid on the joint payoff of $\bar \$ _A (p, p_1, q, q_1) + \bar \$ _B (p, p_1, q, q_1)$, which is given as below:
\begin{equation}\label{JointPayoffGMIII}
\begin{array}{l}
 \bar \$ _A (p,p_1 ,q,q_1 ) + \bar \$ _B (p,p_1 ,q,q_1 ) = ( 2\left| {u_{13} } \right|^2  + 2\left| {u_{23} } \right|^2  + 2\left| {u_{31} } \right|^2  + 2\left| {u_{32} } \right|^2  + 4\left| {u_{33} } \right|^2  \\
 \quad \quad \quad \quad   \quad \quad  + 2p\left| {u_{11} } \right|^2  + 2p\left| {u_{12} } \right|^2  + 2p\left| {u_{13} } \right|^2  - 2p\left| {u_{31} } \right|^2  - 2p\left| {u_{32} } \right|^2  - 2p\left| {u_{33} } \right|^2  \\
 \quad \quad \quad \quad   \quad \quad  + 2q\left| {u_{11} } \right|^2  - 2q\left| {u_{13} } \right|^2  + 2q\left| {u_{21} } \right|^2  - 2q\left| {u_{23} } \right|^2  + 2q\left| {u_{31} } \right|^2  - 2q\left| {u_{33} } \right|^2 ) / d \\
 \end{array}
\end{equation}

Equation (\ref{JointPayoffGMIII}) is relevant with $p$, $q$, $u_{ij}$, $i,j=1,2,3$, whose maximum value can be found by the following optimization problem.
\begin{equation}
\begin{array}{l}
 \max \quad \bar \$ _A (p,p_1 ,q,q_1 ) + \bar \$ _B (p,p_1 ,q,q_1 ) \\
 s.t.\quad \left\{ \begin{array}{l}
 \sum\limits_{i,j = 1,2,3} {\left| {u_{ij} } \right|^2 }  = 1 \\
 p,q \in [0,1] \\
 \end{array} \right. \\
 \end{array}
\end{equation}

It is easy to find that the maximum value of $\bar \$ _A (p,p_1 ,q,q_1 ) + \bar \$ _B (p,p_1 ,q,q_1 )$ is $4/d$ where $|u_{i^*j^*}|^2 = 1$, $\forall i \ne i^* \; \rm{and} \; j \ne j^*$, $|u_{ij}|^2 = 0$. Recalling the approach of MW scheme, we know that $\left| {\psi _{in} } \right\rangle  = \left| {ij} \right\rangle$ means the quantum game degrades into a classical game. The result implies two points: (i) The quantum version of GM III can not produce better result -- bigger joint payoff -- compared with the classical version; (ii) In a NE of quantum GM III, the outcomes may be not zero-sum. Let us review NEs in the classical GM III: a NE with joint payoff 0 when $d \ge 1/(b+c)$, and a NE with joint payoff $4/d$ when $d \le 1/(b+c)$. So, in the classical GM III, it is conditional to reach a non-zero-sum NE. If $d > 1/(b+c)$, the GM III still produce zero-sum outcomes.

However, the situation is different in quantum game of GM III. It is possible to find a NE which realizes a non-zero-sum equilibrium but without the condition $d \le 1/(b+c)$. For example, assuming $|u_{11}|^2 = |u_{33}|^2 = 0.5$, namely the initial state is $\left| {\psi _{in} } \right\rangle  = \sqrt{0.5}\left| {11} \right\rangle  + \sqrt{0.5}\left| {33} \right\rangle$. Based on the generalized MW quantum scheme, the expected payoffs of two players can be obtained:
\begin{equation}
\begin{array}{l}
 \bar \$ _A (p,p_1 ,q,q_1 ) = (adp_1  - adq_1  + bdp_1  - bdq_1  + 2)/2d \\
 \bar \$ _B (p,p_1 ,q,q_1 ) = (adq_1  - adp_1  - bdp_1  + bdq_1  + 2)/2d \\
 \end{array}
\end{equation}
A NE $(p^*,p_1^* ,q^*,q_1^* )$ can be found by imposing the following conditions:
\begin{equation}\label{QuantumGMIIINE1}
\begin{array}{l}
 \bar \$ _A (p^* ,p_1^* ,q^* ,q_1^* ) - \bar \$ _A (p,p_1 ,q^* ,q_1^* ) = (p_1^*  - p_1 )(a + b)/2 \ge 0,\forall p_1  \in [0,1] \\
 \bar \$ _B (p^* ,p_1^* ,q^* ,q_1^* ) - \bar \$ _B (p^* ,p_1^* ,q,q_1 ) = (q_1^*  - q_1 )(a + b)/2 \ge 0,\forall q_1  \in [0,1] \\
 \end{array}
\end{equation}
Obviously, inequations (\ref{QuantumGMIIINE1}) are satisfied if $p_1^* = q_1^* = 1$. Therefore, a set of NEs, $\{ (p^* ,p_1^* ,q^* ,q_1^* ) \; | \; p^* ,q^*  \in [0,1],p_1^*  = 1,q_1^*  = 1\}$, is found. At these equilibrium points, the joint payoff of $A$ and $B$ is $2/d$, and $\bar \$ _A (p^* ,p_1^* ,q^* ,q_1^* ) = 1/d$ and $\bar \$ _B (p,p_1 ,q,q_1 ) = 1/d$. More importantly, these NEs do exist unconditionally, given such an initial state. So, the results show that the quantization to GM III produces win-win outcomes without the need of any specific conditions. Here, the quantization changes fascinatingly the properties of GM III, which brings new insight on the opinion formation.

\section{Conclusion}\label{SectConclusion}
In this paper, the quantization is first used in opinion formation games based on the MW quantum scheme. We have investigated three basic game models of opinion formation. Some interesting results have been revealed. For these opinion formation game models which are zero-sum, the quantum versions of these games preserve the zero-sum nature. However, for the game model which would conditionally produce non-zero-sum outcomes, for example the model considering the distance between two individuals with different opinions, the quantization generates win-win outcomes with removing the original condition required. This work brings new insight to opinion formation and evolution. In the future research, we will further study more realistic opinion game models, such as ``Stubborn individuals and Orators" (SO) model \cite{di2007opinion} which has considered the diversity of characters and behaviors of individuals, by using quantization approaches.

\section*{Acknowledgments}
The work is partially supported by China Scholarship Council, National Natural Science Foundation of China (Grant No. 61174022), Specialized Research Fund for the Doctoral Program of Higher Education (Grant No. 20131102130002), R\&D Program of China (2012BAH07B01), National High Technology Research and Development Program of China (863 Program) (Grant No. 2013AA013801), the open funding project of State Key Laboratory of Virtual Reality Technology and Systems, Beihang University (Grant No.BUAA-VR-14KF-02).

%



\bibliographystyle{elsarticle-num}
\bibliography{References}







\end{document}